\documentclass[a4paper,fleqn,usenatbib]{mnras}

\usepackage[T1]{fontenc}
\usepackage{ae,aecompl}
\usepackage{pifont}

\usepackage{graphicx}	
\usepackage{amsmath}	
\usepackage{amssymb}	

\title[LMC GCs]{Accurate radial velocity and metallicity of the Large
Magellanic Cloud old globular clusters NGC\,1928 and NGC\,1939}

\author[Piatti et al.]{
A.E. Piatti$^{1,2}$\thanks{E-mail: andres@oac.unc.edu.ar}, N. Hwang$^{3}$,
A.A. Cole$^{4}$, M.S. Angelo$^{5}$ and B. Emptage$^{4}$
\\
$^{1}$Consejo Nacional de Investigaciones Cient\'{\i}ficas y T\'ecnicas, Av. Rivadavia 1917, 
C1033AAJ, Buenos Aires, Argentina\\
$^{2}$Observatorio Astron\'omico de C\'ordoba, Laprida 854, 5000, 
C\'ordoba, Argentina\\
$^{3}$Korea Astronomy and Space Science Institute, 776 Daedeokdae-Ro Yuseong-Gu, Daejeon 305-348, Korea \\
$^{4}$School of Natural Sciences, University of Tasmania, Private Bag 37, Hobart, 
7001 TAS, Australia\\
$^{5}$Laborat\'orio Nacional de Astrof\'{\i}sica, R. Estados Unidos 154, 37530-000 Itajub\'a, MG, Brazil\\
}

\date{Accepted XXX. Received YYY; in original form ZZZ}

\pubyear{2018}

\begin{document}
\label{firstpage}
\pagerange{\pageref{firstpage}--\pageref{lastpage}}
\maketitle

\begin{abstract}
We present results obtained from spectroscopic observations of red giants located in 
the  fields of the Large Magellanic Cloud (LMC) globular clusters (GCs) NGC\,1928 and NGC\,1939.
We used the GMOS and AAOmega+2dF spectrographs to obtain spectra centred on the 
Ca\,II triplet, from which we derived individual radial velocities (RVs) and 
metallicities. From cluster members we derived
mean RVs of RV$_{\rm NGC\,1928}=249.58\pm$4.65 km/s and 
RV$_{\rm NGC\,1939}=258.85\pm$2.08 km/s,
and mean metallicities of [Fe/H]$_{\rm NGC\,1928}=-1.30\pm$0.15 dex and  
[Fe/H]$_{\rm NGC\,1939}=-2.00\pm$0.15 dex. We found that both GCs have
RVs and positions consistent with being part of the LMC disc, so that we rule
out any possible origin but that in the same galaxy. By computing the best
solution of a disc that fully contains each GC, we obtained circular velocities for 
the 15 known LMC GCs. 
We found that 11/15 of the GCs share the LMC rotation derived 
from $HST$ and $Gaia$ DR2 proper motions. This outcome reveals that the LMC disc
existed since the very early epoch of the galaxy formation and experienced the
steep relatively fast chemical enrichment shown by its GC metallicities.
The four remaining GCs turned out to have circular velocities 
not compatible with an {\it in situ} cluster formation, but rather with being
stripped from the SMC. 

\end{abstract}




\section{Introduction}

Only fifteen old GCs (GCs, ages $\ga$ 12 Gyr) are known to survive in the Large Magellanic 
Cloud (LMC) \citep{pg13}, of which NGC\,1928 and NGC\,1939 have only recently been added 
 by \citet[][hereafter D99]{detal99}. Their first colour-magnitude diagrams come from $HST$ photometry 
\citep{mg04}, confirming their old ages. As far as we are aware, neither NGC\,1928 nor 
NGC\,1939 have published accurate metallicity or radial velocity (RV) measurements.

The orbital motions of LMC ancient GCs are satisfactorily described by a 
disc-like rotation with no GC appearing to have halo kinematics
\citep{shetal10}. \citet{s92} found that these clusters form a disc that
agrees with the parameters of the optical isophotes and inner H\,I rotation curve.
There are some other galaxies that appear to have GC systems with kinematic 
properties related to the H\,I discs \citep[e.g.][]{olsenetal2004}, which might suggest a benign 
evolutionary history, such as might be expected if the LMC has evolved in a low density environment. 

However, the destruction of a GC system that is on a coplanar orbit about a 
larger galaxy could also produce such a disc-like rotation geometry \citep{leamanetal2013}. 
Furthermore, \citet{vdbergh2004} showed that the possibility that the LMC old GCs 
formed in a pressure-supported halo, rather than in a rotating disc, should not be discarded. 
In this sense, \citet{carreraetal2008} argued that the lack of evidence of such a hot 
stellar halo in the LMC is related to a low contrast of the halo population with respect to 
that of the disc, particularly at the innermost galactocentric radii where NGC\,1928 and 
NGC\,1939 are located. On the other hand, \citet{carpinteroetal2013} modelled the dynamical 
interaction between the Small Magellanic Cloud (SMC) and the LMC, and found that at least some
 of the oldest clusters observed in the LMC could have originated in the SMC. 

The LMC old GCs have also been compared to those of the Milky Way (MW).
\citet{brocatoetal1996}, \citet{mucciarellietal2010} and \citet{wagnerkaiseretal2017}, among othes,
showed that the old LMC GCs resemble the MW ones
in age and in many chemical abundance patterns. In contrast, \citet{johnsonetal2006} found that many 
of the abundances in the LMC old GCs are distinct from those observed in the MW, while
\citet{pg13}  suggested that the most likely explanation for the difference between the old GC and field star age-metallcity relationships is a very rapid early chemical enrichment traced 
by the very visible old GCs.
Indeed,  the integrated
spectroscopic metallicities obtained by \citet{detal99} suggest that NGC\,1928 is one of
the most metal-rich ([Fe/H] $\sim$ -1.2 dex) old GCs, whereas NGC\,1939 one 
of the most metal-poor ([Fe/H] $\sim$ -2.0 dex) old GCs.

In Section 2 we describe the spectroscopic observations performed with the aim of
deriving for the first time accurate mean cluster RVs (Section 3) and metallicities
(Section 4). These quantities are considered in Section 5 to investigate whether
NGC\,1928 and NGC\,1939 have been born in the LMC disc, or have other origins.
Finally, a summary of the results is presented in Section 6.

\section{Observational data sets}

\begin{figure*}
   \includegraphics[width=\columnwidth]{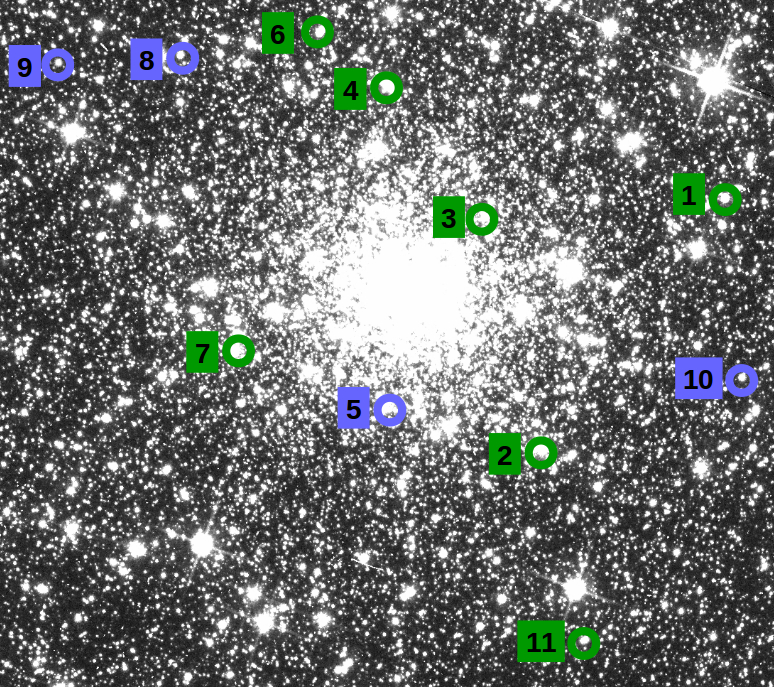}
   \includegraphics[width=\columnwidth]{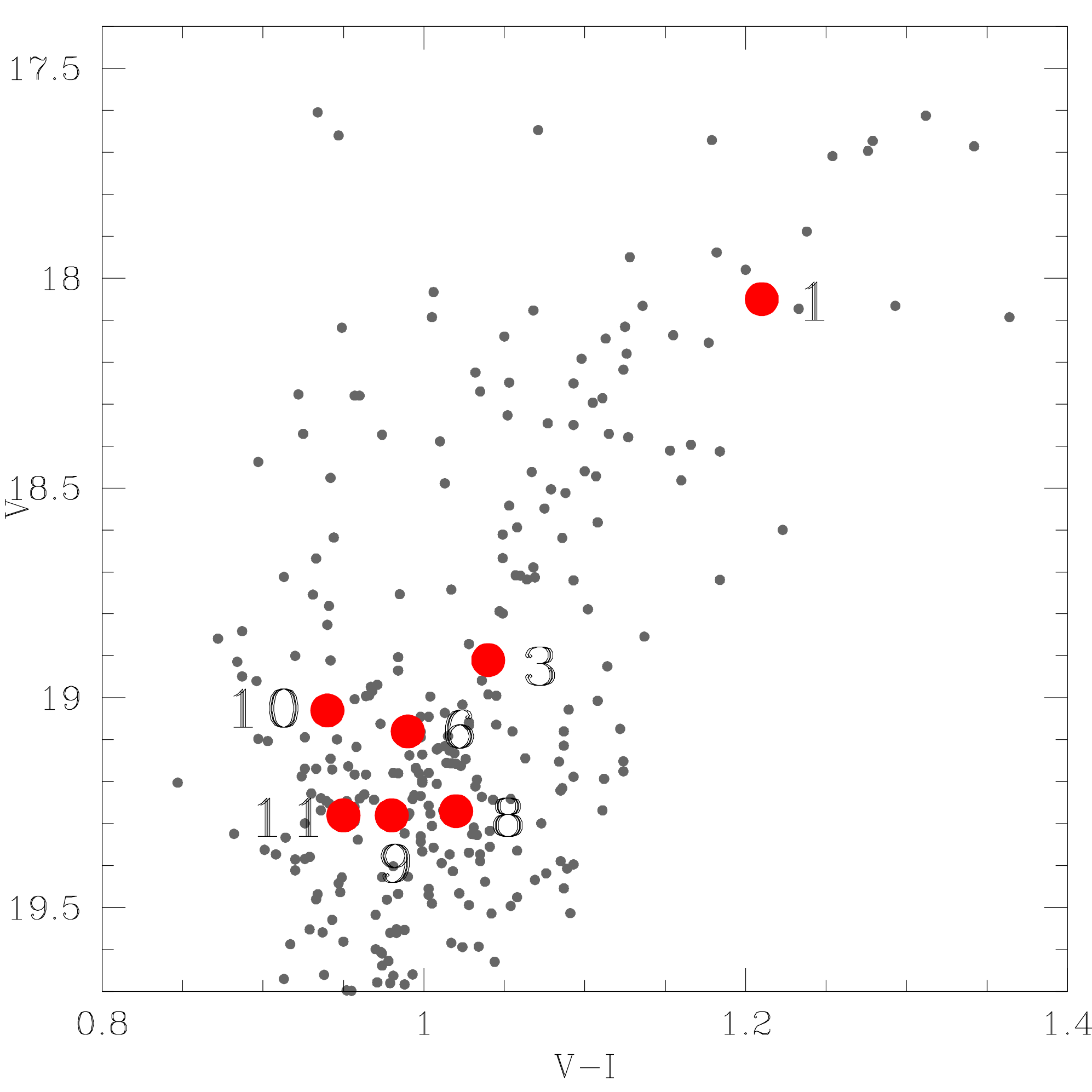}
   \includegraphics[width=\columnwidth]{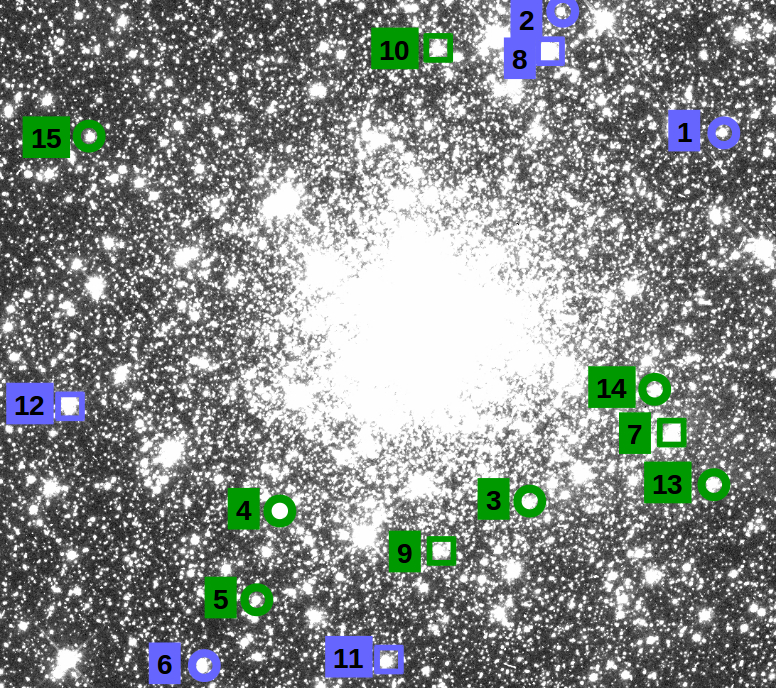}
   \includegraphics[width=\columnwidth]{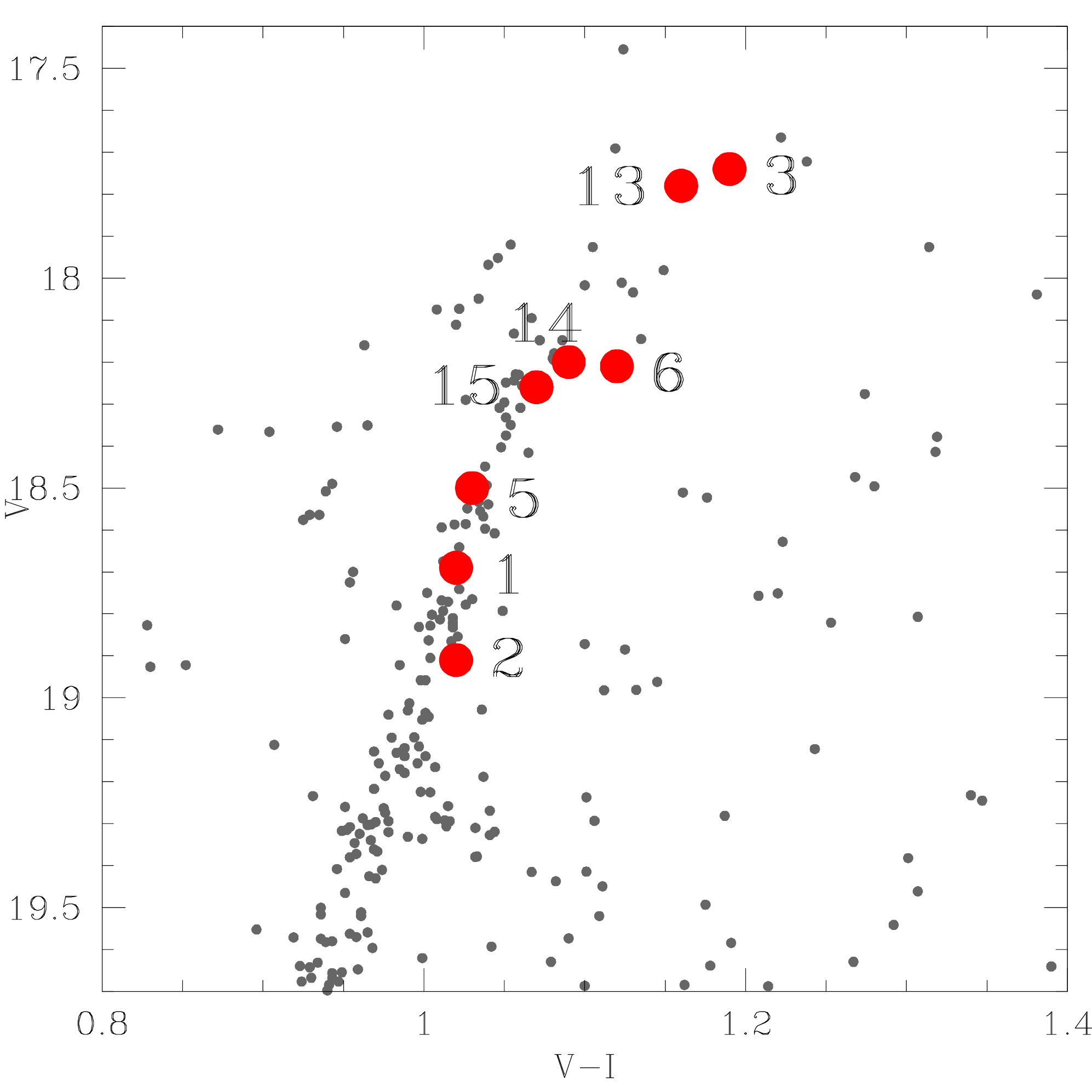}
    \caption{40$\arcsec$$\times$40$\arcsec$ publicly available $F555W$ images centred on 
NGC\,1928 (top) and on NGC\,1939 (bottom) with the selected targets labelled
with the ID numbers of Table\,1. Green and blue symbols represent member and non-member stars,
respectively (see Table~\ref{tab:table2}); while circles and boxes correspond to stars
observed with GMOS and AAOmega+2dF, respectively.
The right-hand panels show the cluster CMDs for the respective image areas, where 
selected stars with available \citet{mg04}'s photometry are higlighted with red 
filled circles.}
   \label{fig:fig1}
\end{figure*}

\begin{figure}
   \includegraphics[width=\columnwidth]{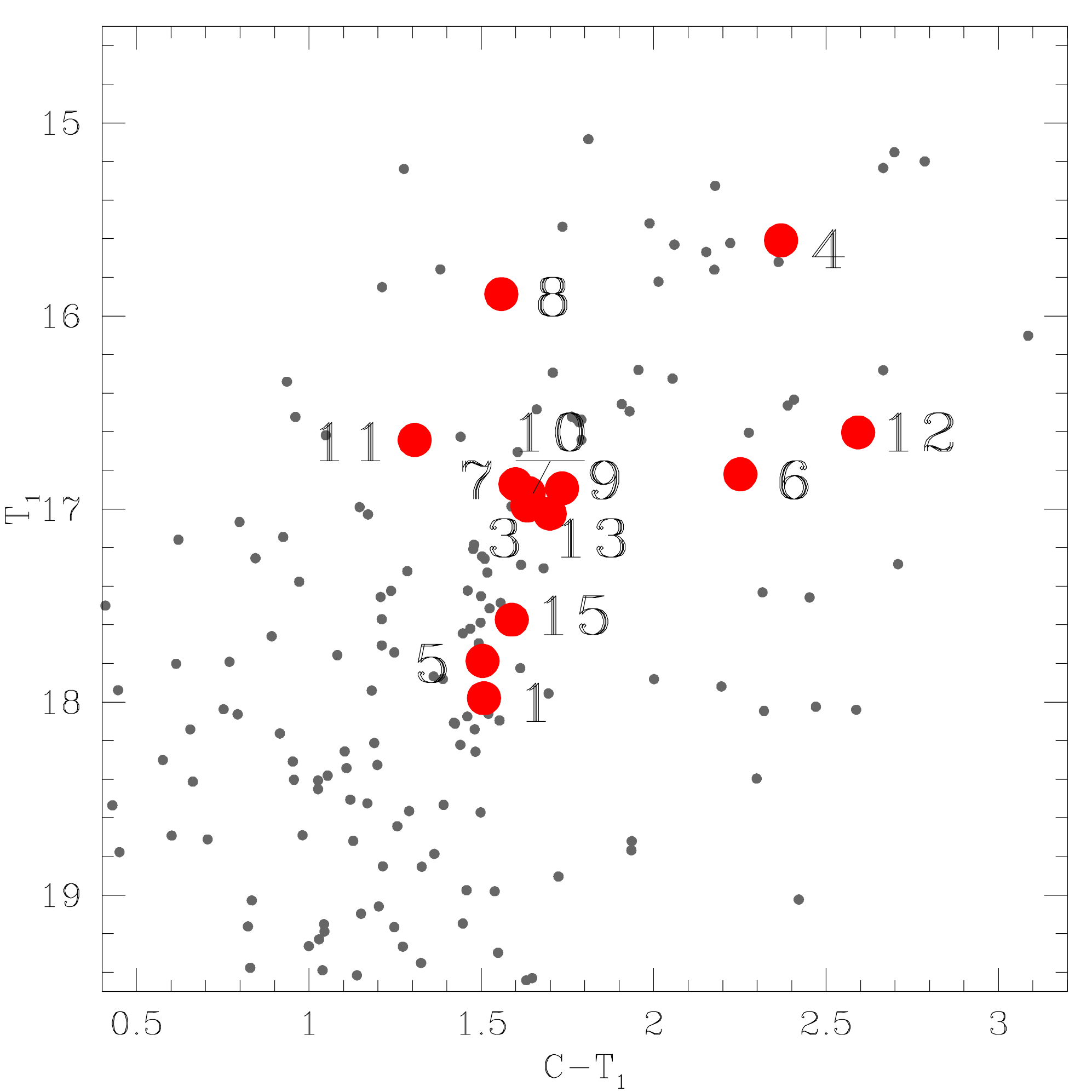}
    \caption{NGC\,1939's $CT_1$ CMD for stars located in the same area as in Fig.~\ref{fig:fig1}, 
taken from \citet{p17e}. Selected stars are highlighted with filled red circles.}
   \label{fig:fig2}
\end{figure}

We carried out spectroscopic observations centred on the Ca\,II infrared triplet 
($\sim$ 8500 \AA) of red giant stars located in the fields of NGC\,1928 and 1939.
Most of the targets were selected from the $HST$ photometric data set of 
\citet{mg04}, bearing in mind their loci in the cluster colour-magnitude
diagrams (CMDs). Because of the relatively small cluster
angular sizes ($\la$ 1 arcmin) and their high crowding, many
cluster red giants were discarded. For this reason, we considered some few other relatively 
bright red giant stars (4 in NGC\,1928 and 1 in NGC\,1939) without $HST$ photometry.
Fig.~\ref{fig:fig1} illustrates the positions of the selected targets in the
cluster fields and CMDs, respectively. In the case of NGC\,1939, we have also available 
$CT_1$ Washington photometry \citep{p17e}, from which we built the cluster CMD of 
Fig.~\ref{fig:fig2}.

\subsection{Gemini South Observatory: GMOS spectra}

We carried out spectroscopic observations of stars in the field of NGC\,1928 and 
NGC\,1939 using the 
Gemini Multi-Object Spectrograph (GMOS) of Gemini South observatory during the nights of October 
21 and 25, 2017, through programmes GS-2017B-Q-23 and GS-2017B-Q-71 (PI: Piatti), 
respectively. For each star cluster, we took four consecutive exposures of 900 sec 
for a single mask, as well as CuAr arcs and flats before and after the individual science 
exposures in order to secure a stable wavelength calibration. The total integration time
 for the science targets was 3600 sec. We used the R831 grating and the OG515 (> 520 nm) filter,
 combined with a mask of 1.0 arcsec wide slits placed on the target stars, which gave a
spectral sampling of $\sim$ 0.75\AA\,per pixel with the 2$\times$2 CCD binning configuration. We 
observed 11 and 9 science target stars in the field of NGC\,1928 and NGC\,1939, respectively.

\begin{figure}
   \includegraphics[width=\columnwidth]{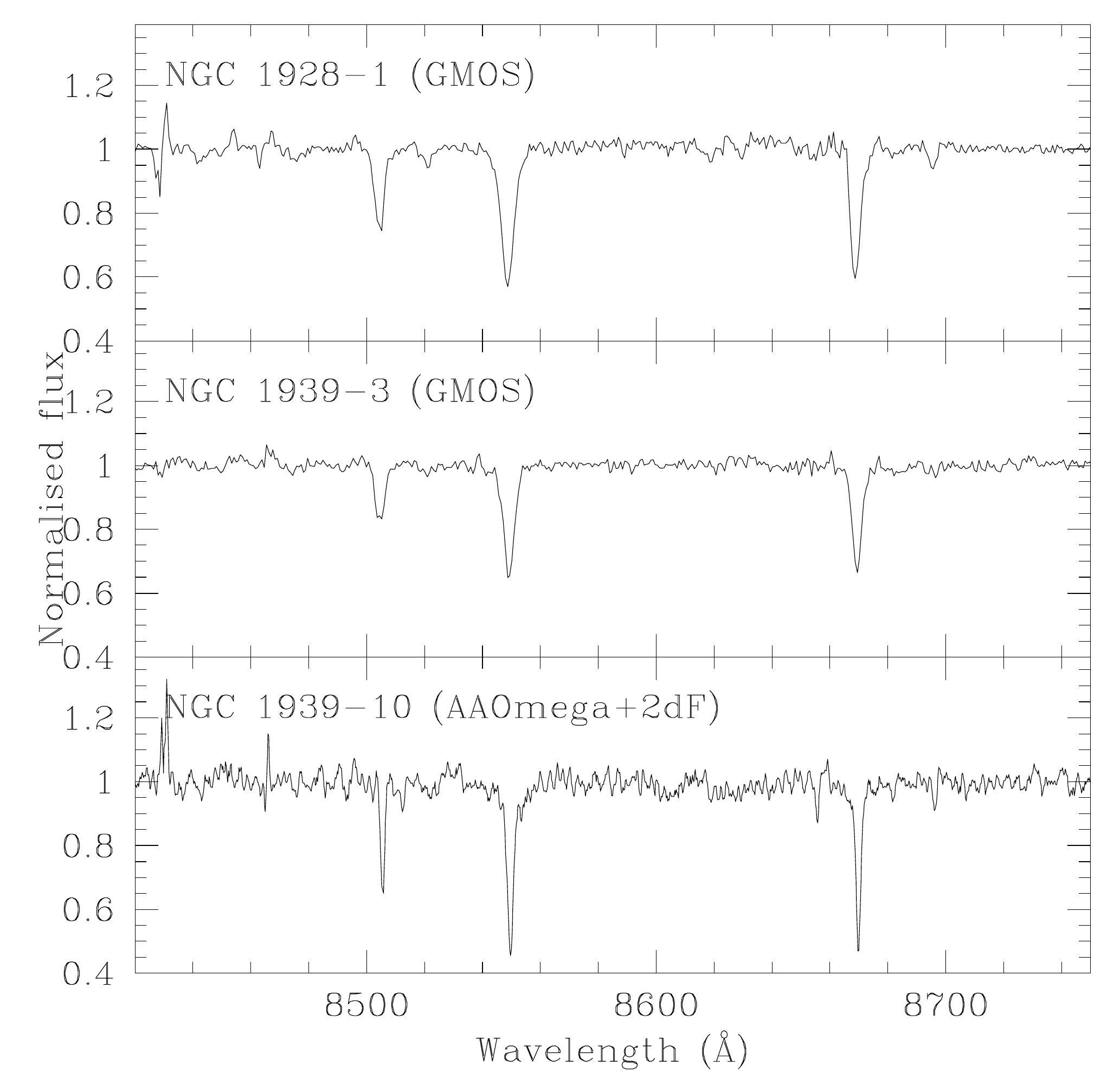}
    \caption{Normalised spectra of some observed stars.}
   \label{fig:fig3}
\end{figure}

We reduced the spectra following the standard GMOS data reduction procedure using the
IRAF.{\sc gemini.gmos} package. The wavelength calibration was derived using the
{\sc gswavelength} task, which compares the observed spectra with GCAL arc lamp data, 
and a wavelength solution was derived with a rms less than 0.20\AA. We also used sky 
OH emission lines to further constrain the wavelength calibration and applied
small offsets of about 0.3$-$0.5\AA\, to the science spectra. The final dispersion of our 
data turned out to be 26.47 km/sec per pixel and the S/N ratio of the resulting spectra ranges 
from 30 up to 100, measured using the local continuum of the Ca\,II triplet.
Fig.~\ref{fig:fig3} illustrates spectra of some science targets.


\subsection{Anglo-Australian Telescope: AAOmega+2dF spectra}

We observed the region around ($\alpha$, $\delta$) = (5:24, $-$68:48) with the AAOmega spectrograph and 2dF fibre positioner at the 3.9m AAT on 2017 December 10--11, as part of
a followup program intended to identify the most metal-poor red giants in the LMC (Emptage et al., in preparation). The fibre positions were chosen to optimise overlap with the targets in 
\citet{coleetal2005} and \citet{vanderswaelmenetal2013} in order to provide metallicity cross-calibration. NGC\,1939 is not far from the field centre, so 6 fibres were assigned to red giants within 3$^{\prime}$ of the cluster, over two configurations of the fibre plate. No fibers were assigned to stars in the vicinity of NGC~1928, a sparser cluster farther from the 2dF field centre. 

On 10 December, the field was observed for 3$\times$1800~s in 1$\farcs$6 seeing, and the following night a second fibre configuration was observed for 3$\times$1200~s in 1$\farcs$4 seeing. The red arm of the spectrograph was employed with the 1700D grating, centred on $\lambda_c$ = 8600 \AA, for a dispersion of $\approx$0.24 \AA\ per pixel, and a resolution R $\approx$11,000, depending on the position of the fibre image on the CCD. Arc and fibre flat exposures were taken immediately prior to each set of three science exposures.

The data were reduced using the standard {\it 2dfdr} data reduction package, which tunes the extraction parameters to optimise the signal to noise, producing  wavelength-calibrated, sky-subtracted spectra. We obtained typical SNR values in the continuum of $\approx$15--50 depending on target I magnitude and fibre centring accuracy. Continuum normalisation was performed using the IRAF task {\it continuum}, with a sixth-order cubic spline fit and rejection of unusually low points, which are assumed to be photospheric lines. The spectra were not flux-calibrated, as we intend only to measure equivalent widths and radial velocities.

\section{Radial velocity measurements}

\subsection{GMOS spectra}

\begin{figure}
    \includegraphics[width=\columnwidth]{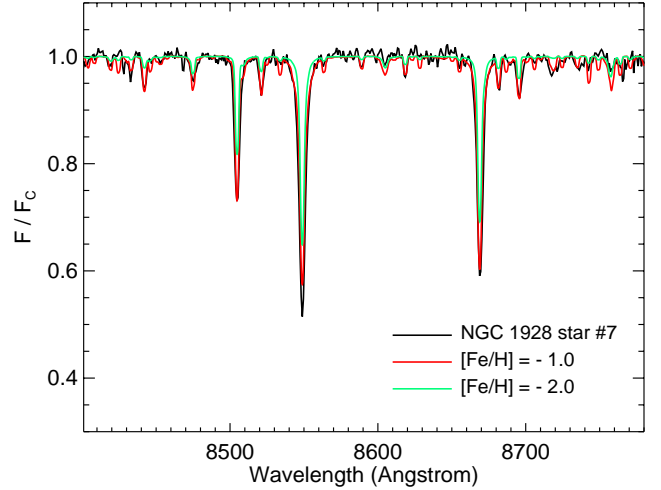}
  \caption{Continuum normalised spectrum of star \#7 of NGC\,1928 (black lines) and 
the best fitted synthetic template spectra for [Fe/H]$=-1.0$\,dex (red line) and $-2.0$\,dex (green line),
respectively.}
  \label{fig:fig4}
\end{figure}

We measured RVs by cross-correlating the observed spectra and synthetic ones 
taken from the PHOENIX library\footnote[1]{http://phoenix.astro.physik.uni-goettingen.de/} 
\citep{husser2013}. The synthetic spectra library covers the wavelength range 
500 $-$ 55000\,\AA\,and 
provides a wide coverage in effective temperatures ($T_{\textrm{eff}}$ $\sim 2300 - 12000\,^{\rm o}$K), 
surface gravities (log\,($g$) 
$\sim 0.0 - 6.0$\,dex) and metallicities ([Fe/H] $\sim -4.0 - +1.0$\,dex). We selected templates 
with $T_{\textrm{eff}}$ in the range $4000-5500\,^{\rm o}$K and log\,($g$) between $1.5-3.0$\,dex, 
which correspond to giant stars with MK types $\sim$ G0$-$K4. In the case of NGC\,1928, we restricted 
the templates to those with [Fe/H]$=-1.0$\,dex, while for NGC\,1939 we employed those with
 [Fe/H]$=-2.0$\,dex (see Section 4). In both cases, we selected 224 templates and checked that 
the restriction in metallicity already has a negligible impact on the RV estimates, since variations 
of 1.0 dex in [Fe/H] resulted in a change of $\sim$ 1 km/s in the derived RV (see also Fig.~\ref{fig:fig4}).

The observed spectra were continuum normalised before the cross-correlation procedure and 
the synthetic templates had their spectral resolution degraded to match the resolution of our
 science spectra. We employed the transformation equations of \citet{ciddor1996} to convert the wavelength 
grids from vacuum ($\lambda_{\textrm{vac}}$) to air wavelengths ($\lambda_{\textrm{air}}$; see also 
Section 3.2 of \citet{angelo2017} for more details). 
Spectral fluxes ($F_{\lambda}^{\textrm{vac}}=\frac{dE_{\lambda}}{dt\,d\lambda_{\textrm{vac}}\,d\textrm{Area}}$) 
were also converted from vacuum to air values through the expression:

\begin{equation}
  F_{\lambda}^{\textrm{air}}=\frac{dE_{\lambda}}{dt\,d\lambda_{\textrm{air}}\,d\textrm{Area}}=F_{\lambda}^{\textrm{vac}}\left(  \frac{d\lambda_{\textrm{vac}}}{d\lambda_{\textrm{air}}}  \right)
\end{equation}

Each observed spectrum was cross-correlated against the whole selected synthetic template sample by making use of
the IRAF.{\sc fxcor} task, which implements the algorithm described in \citet{tonry1979} for the 
construction of the cross-correlation function (CCF) of each object - template pair of spectra. Besides the 
RV estimates, {\sc fxcor} returns the CCF normalised peak ($h$) $-$ an indicator of the degree of similarity 
between the correlated spectra $-$ and the Tonry \& Davis ratio (TDR) 
defined as TDR$=h/(\sqrt{2}\sigma_{a})$, where $\sigma_{a}$ is root mean square of the CCF antisymmetric component. 

For each object spectrum we assigned the RV value resulting from the cross-correlation 
with the highest $h$ value, which was in all cases greater than 0.8. We finally carried 
out the respective heliocentric corrections. 
Table~\ref{tab:table1} lists the resulting RVs with their respective uncertainties, while Fig.~\ref{fig:fig4}
illustrates the cross-correlation procedure. 

\subsection{AAOmega+2dF spectra}

We measured radial velocities by Fourier cross-correlation between the target star spectra and a set of templates obtained at the same resolution and signal-to-noise ratio. The templates were LMC field red giants observed by \citet{coleetal2005}, with metallicities from $-1 \lesssim$ [M/H] $\lesssim$ $-2$, and radial velocities between 200--300~km/s. Relative velocities for each star compared to each template in turn were calculated using {\it fxcor} in IRAF, and converted to a heliocentric frame. The radial velocities based on each template were averaged together, weighted by the cross-correlation peak height. The uncertainty in the resulting average velocities is on the order of $\pm$3~km/s, and is dominated by scatter between the various templates. Therefore the line of sight velocity dispersion of the field stars is highly resolved by these measurements, but cluster is expected to be unresolved.
 
\begin{table*}
\caption{Positions, radial velocities, Ca\,II triplet lines equivalent widths and
metallicities of the selected targets.} 
\label{tab:table1}
\begin{tabular}{@{}lccccccccc}\hline
ID         &Instrument  &  R.A.   &   Dec.  & S/N  &    RV      &  W8498  &  W8542  &   W8662  & [Fe/H] \\
           &  &  (deg)  &  (deg)  &      &  (km/s)    &   (\AA) &  (\AA)  &   (\AA)  &  (dex) \\
\hline
NGC\,1928-1 & GMOS & 80.217297 & -69.475896 & 71.3 & 250.29$\pm$3.03 & 1.024$\pm$0.010 & 2.516$\pm$0.086 & 1.724$\pm$0.023& -1.35$\pm$0.13 \\
NGC\,1928-2 &GMOS & 80.230389 & -69.481931 &102.5 & 245.02$\pm$2.77 & 1.259$\pm$0.090 & 2.606$\pm$0.035 & 1.960$\pm$0.063& ---\\
NGC\,1928-3 & GMOS& 80.234166 & -69.476247 & 33.1 & 238.61$\pm$6.92 & 0.983$\pm$0.040 & 1.963$\pm$0.070 & 1.847$\pm$0.093& -1.32$\pm$0.17 \\
NGC\,1928-4 &GMOS & 80.240765 & -69.472867 & 66.7 & 268.80$\pm$2.52 & 1.832$\pm$0.112 & 3.567$\pm$0.106 & 2.207$\pm$0.107& --- \\
NGC\,1928-5 &GMOS & 80.241292 & -69.480756 & 64.7 & 286.96$\pm$3.37 & 1.863$\pm$0.105 & 3.510$\pm$0.160 & 2.247$\pm$0.048& --- \\
NGC\,1928-6 &GMOS & 80.245312 & -69.471593 & 49.2 & 261.24$\pm$5.92 & 0.816$\pm$0.036 & 2.284$\pm$0.052 & 1.55$\pm$0.056& -1.33$\pm$0.15 \\
NGC\,1928-7 &GMOS & 80.251698 & -69.479367 & 86.3 & 251.72$\pm$3.08 & 1.089$\pm$0.033 & 2.678$\pm$0.071 & 2.039$\pm$0.048& ---\\
NGC\,1928-8 &GMOS & 80.255200 & -69.472097 & 46.3 & 216.78$\pm$6.25 & 1.021$\pm$0.061 & 2.875$\pm$0.139 & 2.329$\pm$0.140& -0.71$\pm$0.24 \\
NGC\,1928-9 &GMOS & 80.263638 & -69.472226 & 53.6 & 214.11$\pm$3.81 & 0.970$\pm$0.032 & 2.780$\pm$0.120 & 1.875$\pm$0.094& -0.93$\pm$0.20 \\
NGC\,1928-10&GMOS & 80.216320 & -69.480298 & 47.7 & 280.37$\pm$5.58 & 1.286$\pm$0.080 & 3.065$\pm$0.114 & 2.316$\pm$0.084& -0.61$\pm$0.23\\
NGC\,1928-11& GMOS& 80.227749 & -69.486631 & 29.0 & 231.37$\pm$8.27 & 0.950$\pm$0.100 & 1.958$\pm$0.157 & 1.631$\pm$0.040& -1.33$\pm$0.21\\ 
            & &           &            &      &               &                 &                 &                & \\
NGC\,1939-1 &GMOS & 80.339326 & -69.944931 & 56.6 & 279.94$\pm$6.76 & 0.630$\pm$0.095 & 1.351$\pm$0.026 & 1.274$\pm$0.035& -1.95$\pm$0.13\\
NGC\,1939-2 &GMOS & 80.350991 & -69.942040 & 39.6 & 258.18$\pm$4.92 & 1.599$\pm$0.133 & 3.203$\pm$0.220 & 2.486$\pm$0.123& -0.41$\pm$0.30\\
NGC\,1939-3 &GMOS & 80.352639 & -69.954109 & 83.1 & 261.36$\pm$3.69 & 0.688$\pm$0.046 & 1.584$\pm$0.030 & 1.351$\pm$0.042& -2.05$\pm$0.10\\
NGC\,1939-4 &GMOS & 80.370202 & -69.954369 & 22.1 & 260.40$\pm$2.93 & 0.932$\pm$0.013 & 2.139$\pm$0.012 & 1.577$\pm$0.068& -2.00$\pm$0.09\\
NGC\,1939-5 &GMOS & 80.371903 & -69.956597 & 40.5 & 241.33$\pm$13.40 & 0.711$\pm$0.060 & 1.576$\pm$0.044 & 0.899$\pm$0.030& -2.02$\pm$0.12\\
NGC\,1939-6 &GMOS & 80.375680 & -69.958244 & 49.9 & 250.52$\pm$3.92 & 1.589$\pm$0.050 & 3.528$\pm$0.095 & 2.634$\pm$0.051& -0.41$\pm$0.19\\
NGC\,1939-7 &AAO+2dF & 80.342251 & -69.952236 & 25.3 & 259.90$\pm$2.50 & 0.900$\pm$0.100 & 1.176$\pm$0.118 & 1.384$\pm$0.138& -2.14$\pm$0.15\\ 
NGC\,1939-8 & AAO+2dF& 80.351829 & -69.943060 & 26.6 & 282.50$\pm$3.40 & 0.800$\pm$0.100 & 2.348$\pm$0.235 & 2.502$\pm$0.250& -1.57$\pm$0.24\\
NGC\,1939-9 &AAO+2dF& 80.358841 & -69.955288 & 24.9 & 259.90$\pm$2.80 & 0.809$\pm$0.081 & 2.048$\pm$0.205 & 1.131$\pm$0.113& -1.94$\pm$0.21\\
NGC\,1939-10& AAO+2dF& 80.359225 & -69.943083 & 19.2 & 259.60$\pm$2.80 & 0.569$\pm$0.057 & 2.067$\pm$0.207 & 1.224$\pm$0.122& -1.98$\pm$0.20\\
NGC\,1939-11&AAO+2dF & 80.362350 & -69.957974 & 13.8 & 257.90$\pm$3.50 & 1.400$\pm$0.100 & 4.016$\pm$0.402 & 2.416$\pm$0.242& -0.60$\pm$0.35\\
NGC\,1939-12& AAO+2dF& 80.385450 & -69.951819 & 25.0 & 270.20$\pm$2.80 & 1.495$\pm$0.150 & 3.880$\pm$0.388 & 2.950$\pm$0.295& -0.44$\pm$0.42\\
NGC\,1939-13&GMOS & 80.339554 & -69.953529 & 76.8 & 268.59$\pm$3.61 & 0.681$\pm$0.040 & 1.747$\pm$0.046 & 1.445$\pm$0.030& -1.95$\pm$0.11\\
NGC\,1939-14&GMOS & 80.343735 & -69.951279 & 40.3 & 253.30$\pm$11.01& 0.502$\pm$0.078 & 1.573$\pm$0.055 & 1.093$\pm$0.030& -1.85$\pm$0.14\\ 
NGC\,1939-15&GMOS & 80.384301 & -69.945343 & 64.0 & 264.30$\pm$3.19 & 0.450$\pm$0.111 & 1.743$\pm$0.041 & 1.401$\pm$0.033& -1.93$\pm$0.14\\

\hline
\end{tabular}
\end{table*}

\section{Overall metallicity estimates}

\subsection{GMOS spectra}

\begin{figure*}
   \includegraphics[width=\columnwidth]{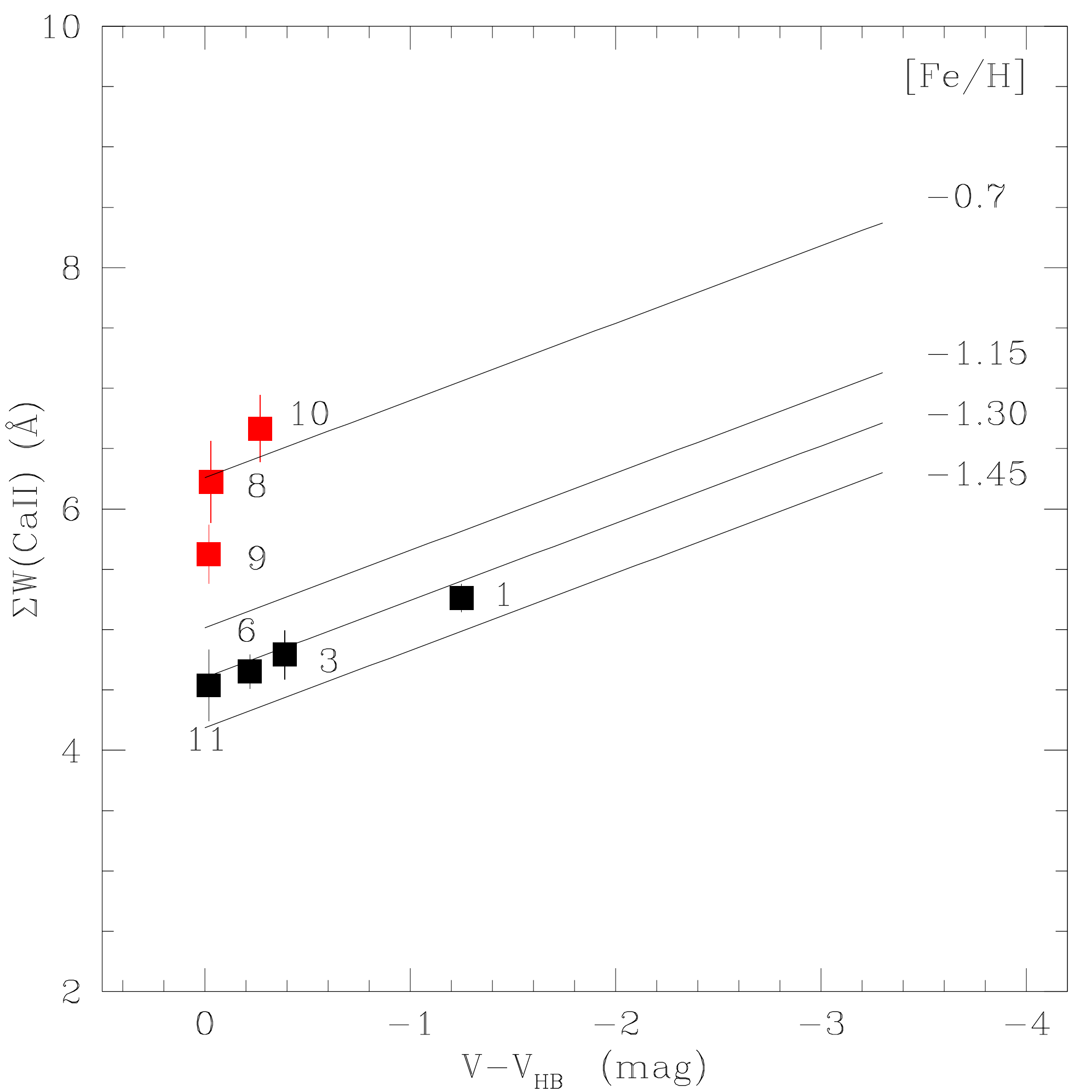}
   \includegraphics[width=\columnwidth]{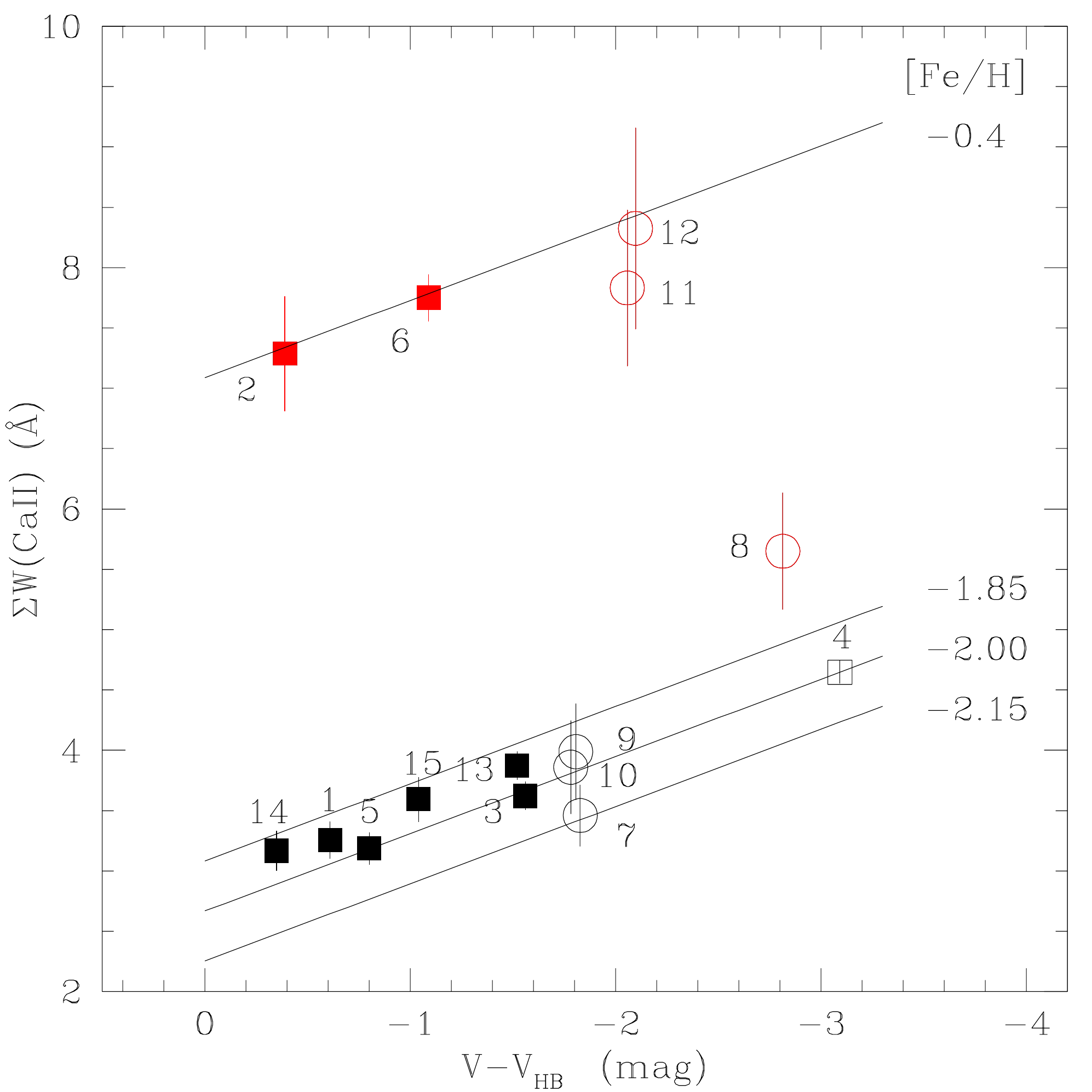}
    \caption{Sum of the CaII triplet line equivalent widths as a function of
$V-V_{HB}$ for stars observed in the fields of NGC\,1928 (left) and NGC\,1939 (right). 
Black and red symbols represent cluster and field stars, respectively,
while boxes and circles correspond to GMOS and AAOmega+2dF spectra, respectively.
Filled and open symbols refers to stars with $HST$ $V$ mag taken from \citet{mg04}
and those with only Washington $T_1$ mags of \citet{p17e} converted into
$V$ ones (see text for details). Errorbars are also drawn. Iso-abundance lines
 derived by \citet{coleetal2004} for some [Fe/H] values are also depicted. 
}
   \label{fig:fig5}
\end{figure*}

Equivalent widths of the CaII infrared triplet lines were measured from the normalised
spectra using the {\sc splot} package within IRAF. Their resulting average values and the 
respective uncertainties are listed in Table~\ref{tab:table1}. The latter were estimated
by computing equivalent widths using different continua, bearing in mind the presence 
of TiO bands and the spectra S/N ratio. We then overplotted the sum
of the equivalent widths of the three CaII lines ($\Sigma$W(CaII)) in the 
$\Sigma$W(CaII) versus $V-V_{\rm HB}$ plane, that has been calibrated in terms of 
metallicity \citep[see, e.g.,][]{coleetal2004}. In that diagram $V_{\rm HB}$ refers
to the mean magnitude of the cluster horizontal branch. For NGC\,1928 and
NGC\,1939 we adopted the individual $V$ magnitudes of the selected stars and   
$V_{\rm HB}$ $=19.3$\,mag, taken from \citet{mg04} (see also Fig.~\ref{fig:fig1}).
We also took advantage of the Washington photometry of \citet{p17e} (see
also Fig.~\ref{fig:fig2}) to convert $T_1$ magnitudes of the selected stars
into $V$ magnitudes $-$ for those stars without $HST$ $V$ mags $-$
using the theoretical red giant branches computed by \citet{betal12}, and the
cluster reddening and distance moduli derived by \citet{mg04}. Fig.~\ref{fig:fig5}
shows the resulting plots, where we included iso-abundance lines according to
eq. (5) of \citet{coleetal2004} for $\beta = 0.64$\,\AA/mag \citep{rutledgeetal97}, while the last
column of Table~\ref{tab:table1} lists the interpolated [Fe/H] values. The errors were
calculated by propagating those of the coefficients in eq. (5) \citep{coleetal2004}, 
$\sigma$($\beta$) \citep{rutledgeetal97}, the $HST$ \citep{mg04} and 
Washington \citep{p17e} photometric errors and $\sigma(\Sigma$W(CaII)),
respectively.
  
\subsection{AAOmega+2dF spectra}

Equivalent widths of the Ca~II triplet lines were measured using the program {\it EW}, originally written by G.S. da~Costa and used by \citet{coleetal2005} and many others \citep[e.g.][]{dacosta2016}. The lines were fit by a sum of Gaussian plus Lorentzian profiles, constrained to have a common centroid. The metallicities were measured as for the GMOS stars, described above. Because of the lower SNR, we tested the results against the method of \citet{starkenburgetal2010}, using only the two strongest lines of the Ca triplet; no significant differences were found within the errorbars. The total error on metallicity is dominated by systematic effects (e.g., possible differences in detailed abundance ratios between the target stars and those used to form the calibration sample) rather than random error from photon noise. For the field stars in the vicinity of NGC\,1939 in common with \citet{coleetal2005}, comparing the equivalent widths measured in the 2017 AAOmega spectra shows an average difference of $\sum$W$_{\mathrm{AAO-VLT}}$ = 0.06 $\pm$0.38~\AA, highly consistent with no systematic offset. 

\section{Analysis and discussion}

We first assigned to the observed stars cluster membership probabilities according to three
different criteria, namely: the position of the stars in the cluster CMDs,
the dispersion of their RV values and that for their [Fe/H] values, respectively. 
For NGC\,1928, we previously discarded stars \#8 and 9, which fall outside the 
cluster radius recently estimated by \citet[31.6$\pm$7.3 arcsec]{pm2018}
from a radial profile that reaches out to $\sim$ 4 times the cluster's
tidal radius. 

By looking at the cluster CMDs (Figs.~\ref{fig:fig1} and \ref{fig:fig2}) we considered 
possible members any star located along the cluster red giant branches, within the 
observed spread of those sequences. We included the results of our assessment in column 
2 of Table~\ref{tab:table2}. Note that this criterion could lead us to conclude on 
the cluster membership of any star that belongs to the LMC star field, because of the
superposition with LMC field features. This is
the case, for instance, of the LMC field red clump.

\begin{figure*}
   \includegraphics[width=\columnwidth]{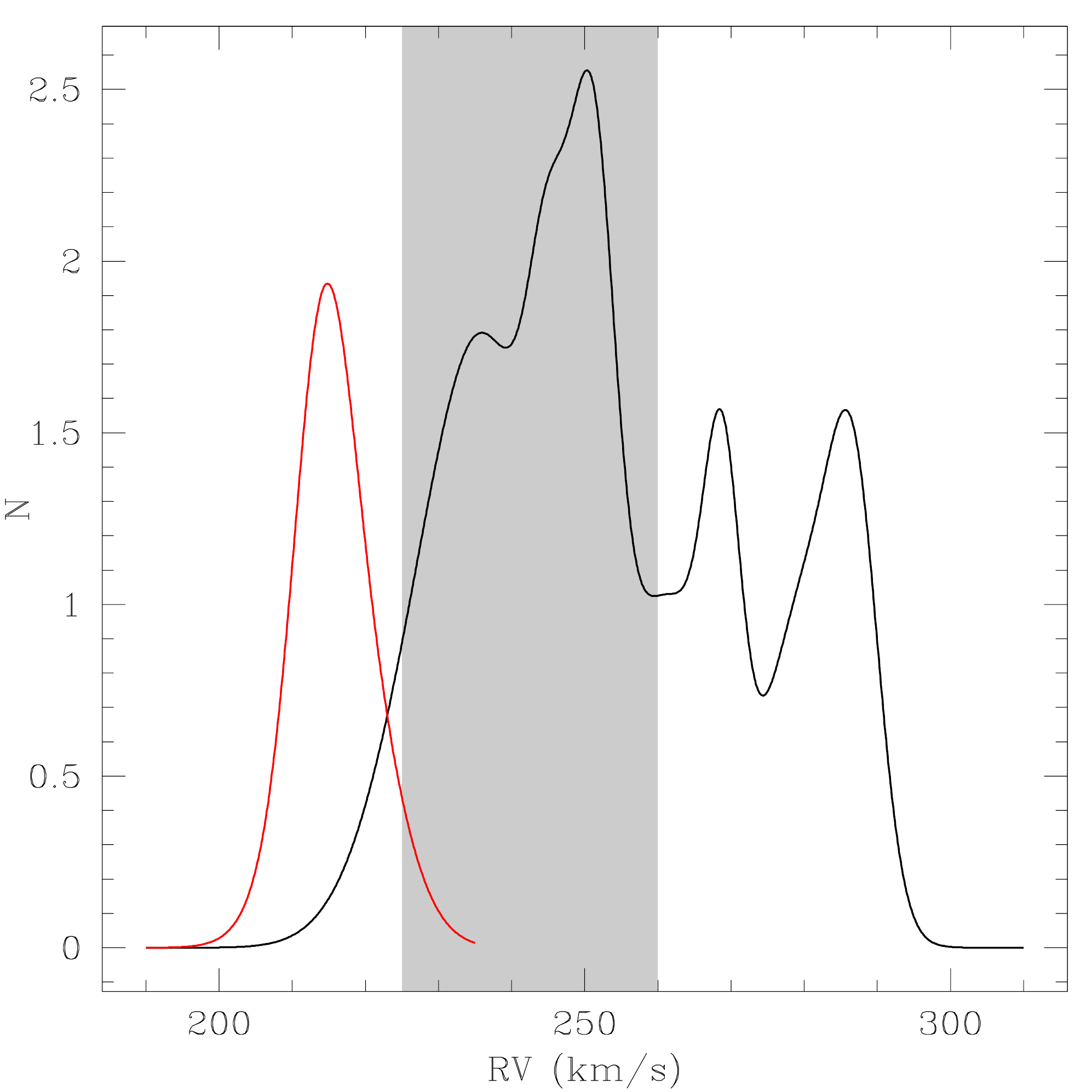}
   \includegraphics[width=\columnwidth]{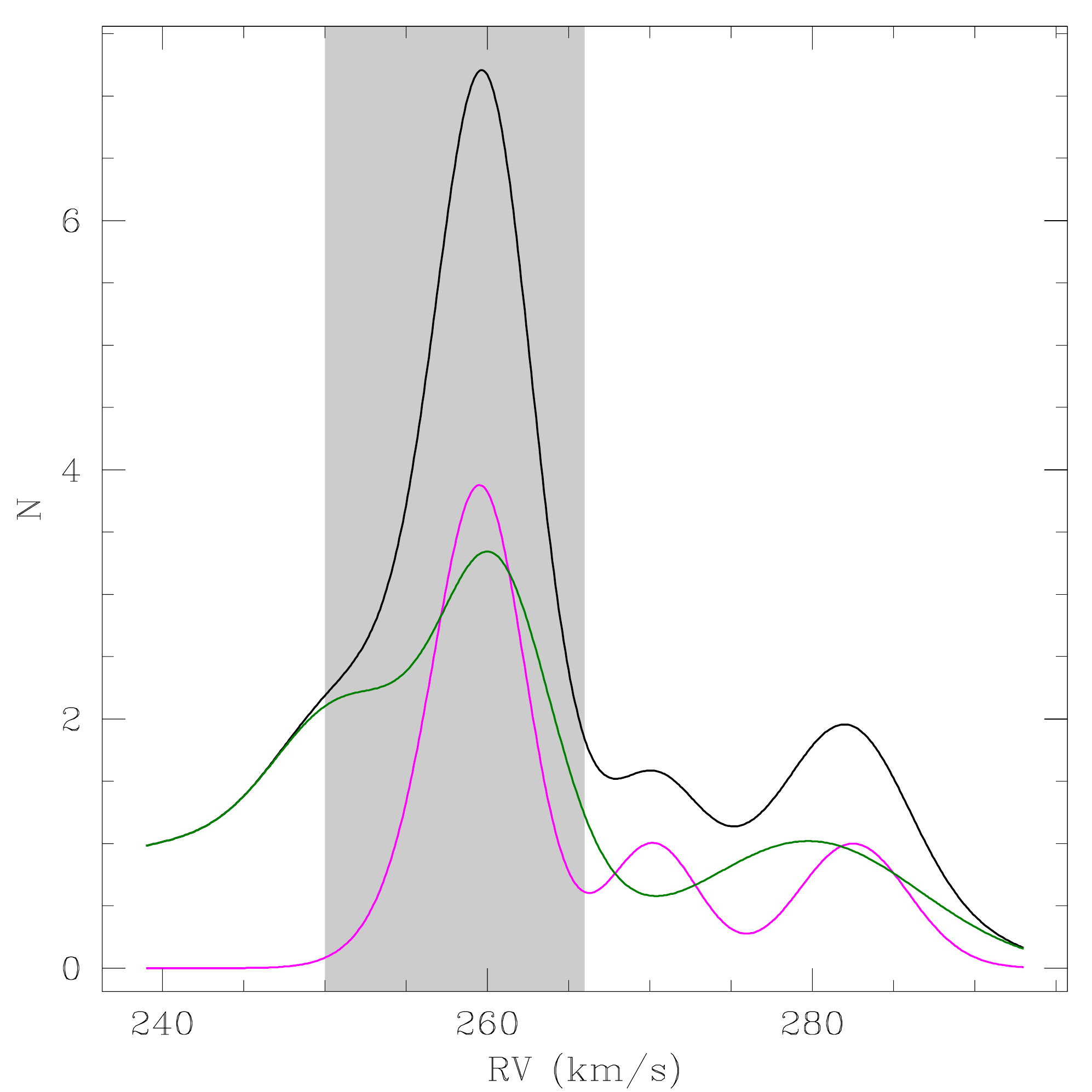}
    \caption{RV distribution functions for stars observed in the field of
NGC\,1928 (left) and NGC\,1939 (right) are drawn with a solid black line.
The red line in the left panel corresponds to the RV distribution of non-member
stars \#8 and 9, while the green and magenta lines in the right panel
refer to the RV distributions obtained from stars observed with GMOS
and  AAOmega+2dF, respectively. The shadowed regions correspond to the cluster
RV ranges.}
   \label{fig:fig6}
\end{figure*}

We then built RV distribution functions by summing all the individual RV values,
each of them represented by a Gaussian with centre and $\sigma$ equal to the 
mean RV value and the associated error, respectively (see Table~\ref{tab:table1}). 
Every Gaussian was assigned the same amplitude. The resulting RV distributions are shown in 
Fig.~\ref{fig:fig6}, where the cluster RV ranges can be clearly identified from the
FWHM of the primary peak (shadowed regions).
In NGC\,1928's panel, we intentionally included stars \#8 and 9 (red curve), thus confirming 
that they are probably non-members. 
In NGC\,1939, we also plotted the RV distributions obtained from using only stars observed with
GMOS (green curve) and with AAOmega+2dF (magenta curve), respectively. As can be seen, there 
is a negligible shift between both RV scales, so that we summed them to produce the overall
RV distribution (black curve). Table~\ref{tab:table2} lists the RV membership status
assigned to each star on the basis of whether its  RV falls within the shadowed regions.

As for the metallicity membership probability, we visually inspected Fig.~\ref{fig:fig5},
in which star sequences along a constant [Fe/H] value can be recognised, with some dispersion.
For instance, the $\Sigma$W(CaII) versus $V-V_{\rm HB}$ diagram for NGC\,1928 (left panel)
shows stars \#8 and 9 $-$ initially discarded because they fall outside the cluster radius $-$
and \#10 at a very distinguishable higher metallicity level. These stars have also RVs quite
different from those of the observed cluster members. For the remaining stars, we do not have any 
argument as to deny them  cluster membership. In the case of NGC\,1939 (right panel), the 
observed more metal-poor sequence contains more than three times the number of
stars in the more metal-rich sequence ([Fe/H] $\sim -0.4$\,dex), so that we concluded that the former 
corresponds to that of the cluster. 
Note that the separation between both sequences is similar for $\Sigma$W(CaII) obtained from
GMOS and AAOmega+2dF spectra, respectively. According to \citet[][see their figure 6]{coleetal2005},  
the derived [Fe/H] values for the observed stars meant to be LMC field stars (red symbols) 
are in excellent agreement with the bulk of metallicity values of LMC bar field giants. 

The final membership status of each star is listed in the last column of Table~\ref{tab:table2}.
Only stars \#1 and 6 observed in the field of NGC\,1939 have RV memberships different from those
adopted using separately their positions in the cluster CMDs and their metallicities, respectively. 
Nevertheless, we rely on the possibility that LMC field stars can have either RVs or metal-contents 
similar to that of the cluster. This is not the case of the field giant \#2 observed also along the 
line-of-sight of NGC\,1939, whose $V$ magnitude and $V-I$ colour place it superimposed on 
the cluster red giant branch (see Fig.~\ref{fig:fig1}). For the remaining stars observed in both
cluster fields, the three membership criteria totally agree. 

\begin{figure}
   \includegraphics[width=\columnwidth]{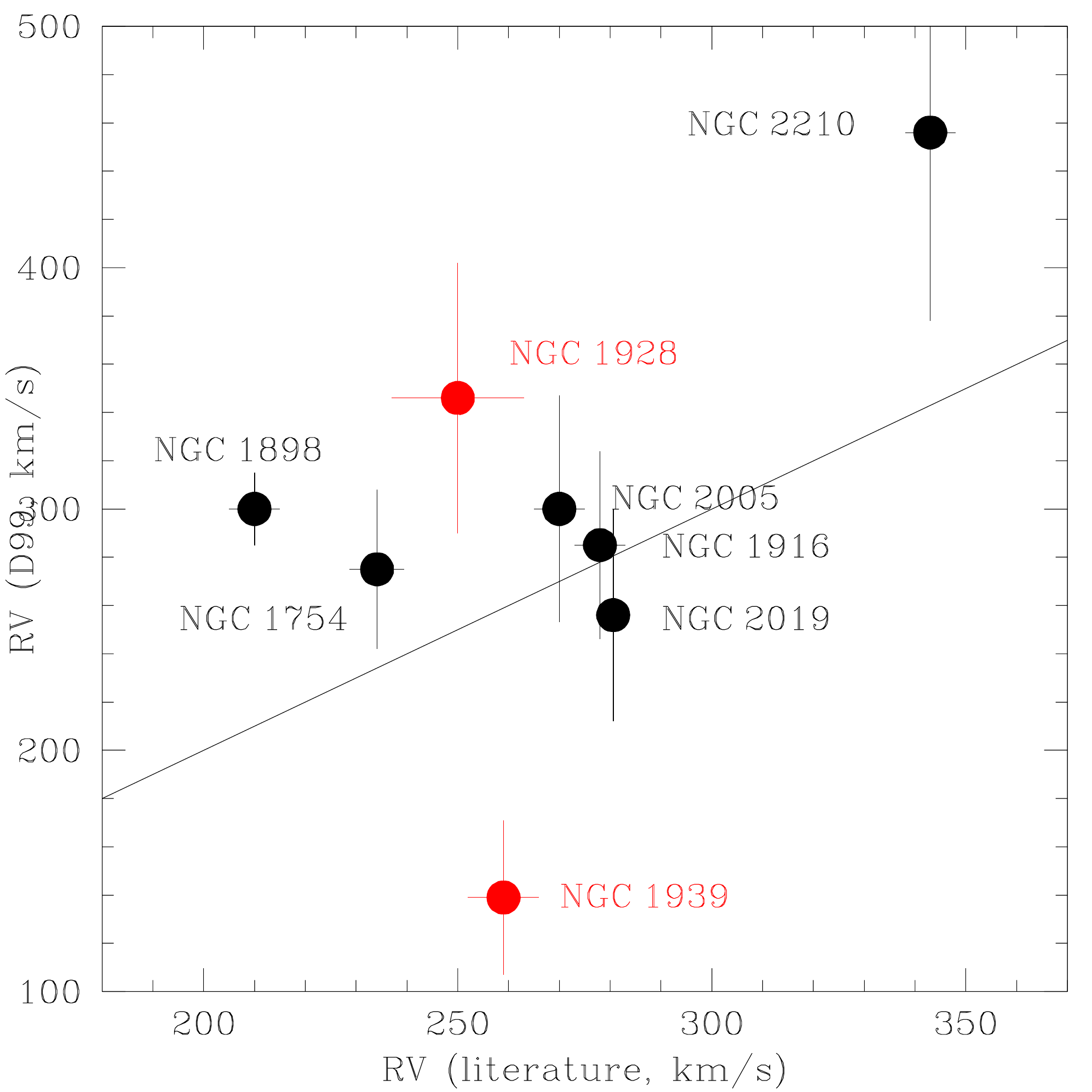}
    \caption{Comparison of LMC GC RVs derived by D99 with
those available in the literature (see Table~\ref{tab:table3}).}
   \label{fig:fig7}
\end{figure}

We finally used the RV and [Fe/H] values of all cluster members to derive the mean cluster
RVs and metallicities by employing a maximum likelihood approach.
The relevance lies in accounting for individual star measurements, which could 
artificially inflate the dispersion if ignored.
We optimized the probability $\mathcal{L}$ that a
given ensemble of stars with velocities RV$_i$ and errors $\sigma_i$ are drawn from a 
population with mean RV $<$$RV$$>$ and dispersion W  
\citep[e.g.,][]{pm1993,walker2006},
as follows:

\begin{equation}
\small
\mathcal{L}\,=\,\prod_{i=1}^N\,\left( \, 2\pi\,(\sigma_i^2 + W^2 \, ) 
\right)^{-\frac{1}{2}}\,\exp \left(-\frac{(RV_i \,- <RV>)^2}{\sigma_i^2 + W^2}
\right) 
.\end{equation}

\noindent where the errors on the mean and dispersion were computed from the respective covariance
matrices\footnote{\citet{pm1993} noted that this approach underestimates the true velocity dispersion
for small sample sizes.}.
We obtained for NGC\,1928, $<$RV$_{\rm NGC\,1928}> = 249.58\pm$4.65
km/s and $<$[Fe/H]$_{\rm NGC\,1928}> = -1.30\pm$0.15 dex, while for NGC\,1939 the mean 
values turned out to be $<$RV$_{\rm NGC\,1939}> = 258.85\pm$2.08 km/s and 
$<$[Fe/H]$_{\rm NGC\,1939}>  = -2.00\pm$0.15 dex. We compared our mean  cluster RVs with 
those previously obtained by D99, who mentioned that
their integrated spectra  were not particularly suitable for accurate velocity measurements.
Fig.~\ref{fig:fig7} shows the results, where other LMC GCs with RV estimates 
available in the literature were added.

\begin{figure}
   \includegraphics[width=\columnwidth]{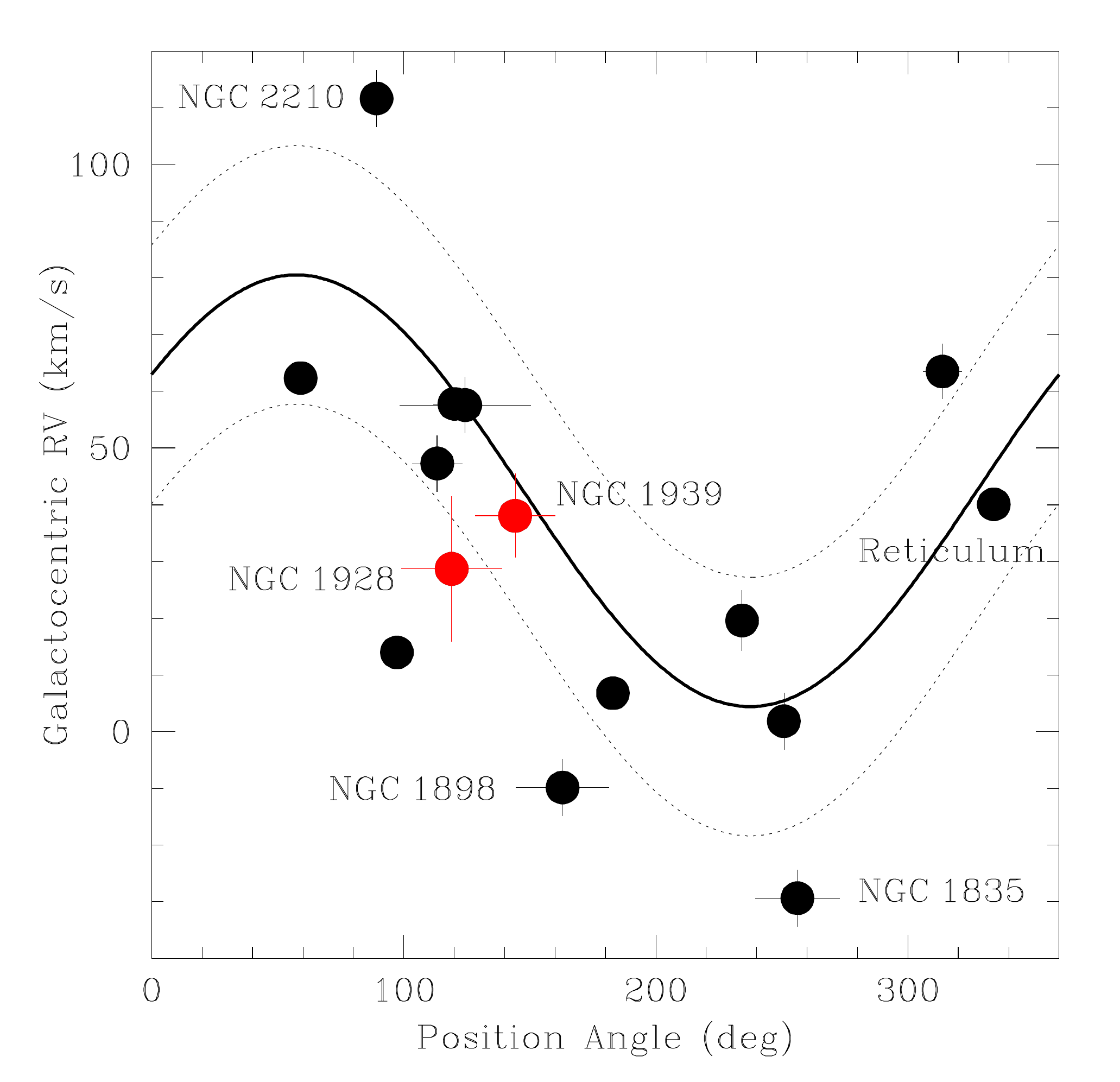}
    \caption{Galactocentric RVs versus PAs diagram for LMC GCs.
RVs were taken from the literature (see Table~\ref{tab:table3}).
We included the curves derived by \citet{vdmk14} 
from $HST$ proper motions of 22 LMC fields (see text for details).}
   \label{fig:fig8}
\end{figure}

One of the diagnostic diagrams most frequently used to assess whether a cluster belongs
to the LMC disc is that which shows the relationship between position angles (PAs) 
and RVs \citep{s92,getal06,shetal10,vdmareletal2002,vdmk14} for a disc-like rotation
geometry. We here followed the recipe used by \citet{s92}, who converted the 
observed heliocentric cluster RVs to Galactocentric RVs through eq.(4) in \citet{fw79}. 
We computed cluster PAs by adopting the LMC disc central coordinates 
and their uncertainties obtained by \citet{vdmk14} from $HST$ average proper motion 
measurements for stars in 22 fields. Fig.~\ref{fig:fig8} shows the disc solution 
derived for those $HST$ proper motions \citep[Table 1 in][]{vdmk14} represented with
a solid line, as well as those 
considering the uncertainties in the LMC disc line-of-sight systemic velocity,  circular 
velocity and  PA of the line-of-nodes and the derived velocity dispersion (dotted lines). 
As can be seen, NGC\,1928 and 1939 are placed 
within the fringes of the LMC disc at 1$\sigma$ confidence, similarly to many of the 
remaining 13 GCs included in the figure for comparison purposes. 

Therefore, assuming that both GCs belong to the LMC disc, we then sought for 
the best disc solutions for their respective RVs and position in the galaxy,
i.e., we looked for the circular velocity ($v_{\rm rot}$) and PA of the line-of-nodes 
(PA$_{\rm LOS}$) of the discs that fully contain them. To do that, we used a grid of 
$v_{\rm rot}$ and PA$_{\rm LOS}$ values to evaluate eq.(1) of \citet{s92} for the cluster 
PAs and their uncertainties, and then to find the most likely pair 
($v_{\rm rot}$,PA$_{\rm LOS}$) that minimizes by $\chi^2$ the difference between the 
cluster RVs with their errors and those calculated above.
We used a grid of $v_{\rm rot}$ from 0.0 up to 200.0 km/s in steps of 1.0 km/s, 
and a range of PA$_{\rm LOS}$ from 0.0 up to 360.0 degrees in steps of 1.0 degree.
For NGC\,1928, the most suitable disc turned out to be that with
$v_{\rm rot} =  45.0\pm$10.0 km/s and PA$_{\rm LOS} =  85.0\pm$10.0 degrees, while the 
resulting one for NGC\,1939 is that with $v_{\rm rot} =  35.0\pm$10.0 km/s and 
PA$_{\rm LOS} = 130.0\pm$10.0 degrees. For comparison purposes, we also computed
$v_{\rm rot}$ and PA$_{\rm LOS}$ values for the remaining LMC GCs
(see Table~\ref{tab:table3}).

\begin{figure}
   \includegraphics[width=\columnwidth]{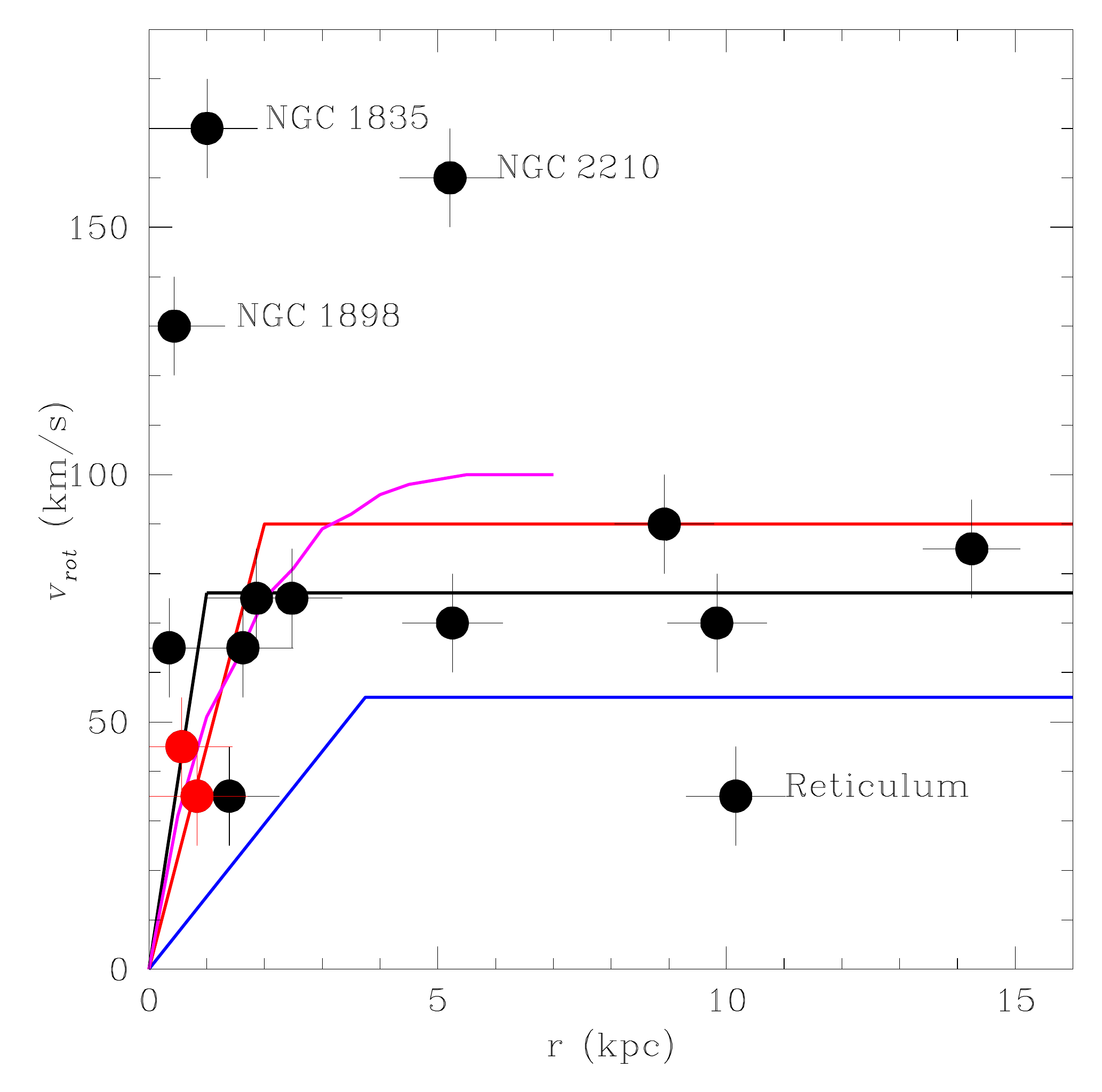}
    \caption{LMC rotation curve as a function of the deprojected distance ($r$)
derived from $HST$ proper motions of 22 fields, and
from LOS velocities of young and old stellar populations drawn with black, red and
blue solid lines, respectively \citep[taken from figure 7 of][]{vdmk14}. The rotation
curve derived by \citet{vasiliev2018} is drawn with a magenta line. All the 15 
GCs have been superimposed with filled circles; NGC\,1928 and 1939 in red.}
   \label{fig:fig9}
\end{figure}

Fig.~\ref{fig:fig9} depicts the resulting $v_{\rm rot}$ values as a function of
the deprojected distances ($r$, see Table~\ref{tab:table3}). The latter were computed using the 
LMC disc fitted  by \citet{vdmk14} from $HST$ proper motions in 22 fields, whose
rotation curve is represented in the figure by a solid black line. The rotation curves 
obtained from line-of-sight (LOS) velocities of young and old stars \citep{vdmk14} are drawn with
red and blue solid lines, respectively, and that from $Gaia$ DR2 proper motions 
\citep{vasiliev2018} with
a magenta line. The figure reveals that NGC\,1928 and
1939 very well match the $HST$ proper motion rotation curve, as also do many
other GCs. Reticulum ($r=$ 10.2 kpc, $v_{\rm rot}=$35 km/s) seems to rotate 
slower than the old stellar population LOS rotation curve, while NGC\,1835, 1898 and 2210 
($v_{\rm rot}>$ 100 km/s) are high circular velocity objects. 

Because the disc-like rotation geometry is shared by most the GCs (age $\ga$ 12 Gyr), 
we infer that the LMC disc had to exist since the early epoch of the galaxy formation,
not only as an structure in itself but also from a dynamical point of view with a non-negligible
angular momentum. The GCs that follow such rotation pattern span the entire
metallicity range of all the GCs in the galaxy (-2.0 $\la$ [Fe/H] (dex) $\la$ -1.3,
see also Table~\ref{tab:table3}), so that the LMC disc had also to experience a similar chemical 
enrichment within $\sim$ 3 Gyr of its GC formation 
\citep[12 $\la$ age (Gyr) $\la$ 14,][]{piattietal2009,wagnerkaiseretal2018}. Furthermore, 
because of the lack of a clear metallicity gradient among the disc GCs, we conclude 
that the whole disc  - except possibly its very outskirts ($r >$ 15 kpc) - has been chemically evolved
similarly.

The four GCs mentioned above that significantly depart from the LMC rotation curve have
ages and metallicities in the same ranges as those disc GCs. However, it is hard
to figure out an {\it in situ} GC scenario for them, because of their very different
$v_{\rm rot}$ values. Note that the velocity dispersion for young and old stellar
population derived by \citet{vdmk14} is 11.6 and 22.8 km/s, respectively 
\citep[see also][]{s92,vdmareletal2002}, so that their velocities differ by more than
three times the LOS velocity dispersion of the LMC old population. One alternative is to 
conclude that these four objects were stripped from the SMC, whose oldest stellar population has
ages and metallicities compatible with them \citep{pg13}. Indeed, such a possibility has been suggested 
by \citet{carpinteroetal2013}, who modelled the dynamical interaction between both galaxies. 
Consequently, our results become in the first observational evidence that the LMC have accreted
not only populations of SMC field stars \citep{olsenetal2011} but also some of its present GCs. 


\begin{table}
\caption{Cluster membership of the observed stars.}
\label{tab:table2}
\begin{tabular}{@{}lccccc}\hline
ID           &  Distance to & CMD   &   RV   & [Fe/H]  & Adopted \\
             & cluster's centre &   &        &         &          \\\hline
NGC\,1928-1  & m &   m    &   m     &   m    &    m     \\
NGC\,1928-2  & m &   --   &   m     &   --   &    m     \\
NGC\,1928-3  & m &    m    &   m     &   m    &    m     \\
NGC\,1928-4  & m &    --   &   m     &   --   &    m     \\
NGC\,1928-5  & m &    --   &   nm    &   --   &    nm    \\
NGC\,1928-6  & m &    m    &   m     &   m    &    m     \\
NGC\,1928-7  & m &    --   &   m     &   --   &    m     \\
NGC\,1928-8  & nm &    --   &   nm    &   nm   &    nm    \\
NGC\,1928-9  & nm &    --  &   nm    &   nm   &    nm    \\
NGC\,1928-10 & m &    m?   &   nm    &   nm   &    nm    \\
NGC\,1928-11 & m &    m    &   m     &   m    &    m    \\
             &         &         &        &          \\
NGC\,1939-1  &  m &   m    &   nm    &   m    &    nm    \\
NGC\,1939-2  &  m &   m    &   nm    &   nm   &    nm    \\
NGC\,1939-3  &  m &   m    &   m     &   m    &    m     \\
NGC\,1939-4  &  m &   m    &   m     &   m    &    m     \\
NGC\,1939-5  &  m &   m    &   m     &   m    &    m     \\
NGC\,1939-6  &  m &   nm   &   m     &   nm   &    nm    \\
NGC\,1939-7  &  m &   m    &   m     &   m    &    m     \\ 
NGC\,1939-8  &  m &   nm   &   nm    &   nm   &    nm    \\
NGC\,1939-9  &  m &   m    &   m     &   m    &    m     \\
NGC\,1939-10 &  m &   m    &   m     &   m    &    m     \\
NGC\,1939-11 &  m &   nm   &   nm    &   nm   &    nm    \\
NGC\,1939-12 &  m &   nm   &   nm    &   nm   &    nm    \\
NGC\,1939-13 &  m &   m    &   m     &   m    &    m     \\
NGC\,1939-14 &  m &   m    &   m     &   m    &    m     \\
NGC\,1939-15 &  m &   m    &   m     &   m    &    m     \\\hline
\end{tabular}
\end{table}

\begin{table*}
\caption{Astrophysical properties of LMC GCs.}
\label{tab:table3}
\begin{tabular}{@{}lcccccccc}\hline
ID          & PA               & $r$      &      RV        & Ref. &   [Fe/H]      & Ref. &  PA$_{\rm LOS}$    & $v_{rot.}$ \\
            & (deg)            & (kpc)    & (km/s)        &       &     (dex)     &      &   (deg)        & (km/s) \\\hline
NGC\,1466 & 250.8$\pm$2.0  & 8.9$\pm$0.9  & 200.0$\pm$5.0 &   1   &-1.90$\pm$0.10 &   10 & 190.0$\pm$10.0 &  90.0$\pm$10.0 \\
NGC\,1754 & 234.2$\pm$6.7  & 2.4$\pm$0.9  & 234.1$\pm$5.4 &   3   &-1.50$\pm$0.10 &  5,6 & 100.0$\pm$10.0 &  75.0$\pm$10.0\\  
NGC\,1786 & 313.7$\pm$7.7  & 1.8$\pm$0.9  & 279.9$\pm$4.9 &   3   &-1.75$\pm$0.10 &  5,6,7& 350.0$\pm$10.0 &  75.0$\pm$10.0 \\
NGC\,1835 & 256.2$\pm$16.8 & 1.0$\pm$0.9  & 188.0$\pm$5.0 &   3   &-1.72$\pm$0.10 &  5,6 & 130.0$\pm$10.0 & 170.0$\pm$10.0 \\
NGC\,1841 & 183.0$\pm$2.0  & 14.2$\pm$0.8 & 210.3$\pm$0.9 &   2   &-2.02$\pm$0.10 &    5 & 130.0$\pm$10.0 &  85.0$\pm$10.0 \\ 
NGC\,1898 & 163.0$\pm$18.5 & 0.4$\pm$0.9  & 210.0$\pm$5.0 &   1   &-1.32$\pm$0.10 & 5,6,8& 110.0$\pm$10.0 & 130.0$\pm$10.0 \\
NGC\,1916 & 124.4$\pm$26.0 & 0.3$\pm$0.9  & 278.0$\pm$5.0 &   1   &-1.54$\pm$0.10 &    9 & 160.0$\pm$10.0 &  65.0$\pm$10.0\\
NGC\,2005 & 113.3$\pm$10.0 & 1.4$\pm$0.9  & 270.0$\pm$5.0 &   1   &-1.74$\pm$0.10 & 5,6,8& 280.0$\pm$10.0 &  35.0$\pm$10.0\\
NGC\,2019 & 120.1$\pm$8.5  & 1.6$\pm$0.9  & 280.6$\pm$2.3 &   2   &-1.56$\pm$0.10 & 5,6,8& 150.0$\pm$10.0 &  65.0$\pm$10.0 \\ 
NGC\,2210 &  89.2$\pm$3.0  & 5.2$\pm$0.9  & 343.0$\pm$5.0 &   1   &-1.55$\pm$0.10 &  7,9 & 140.0$\pm$10.0 & 160.0$\pm$10.0 \\  
NGC\,2257 &  59.1$\pm$1.5  & 9.8$\pm$0.9  & 301.6$\pm$0.8 &   2   &-1.77$\pm$0.10 & 5,7,9& 100.0$\pm$10.0 &  70.0$\pm$10.0 \\
Hodge\,11 &  97.3$\pm$2.8  & 5.2$\pm$0.9  & 245.1$\pm$1.0 &   2   &-2.00$\pm$0.10 &    11& 335.0$\pm$10.0 &  70.0$\pm$10.0 \\  
Reticulum & 334.1$\pm$2.0  & 10.2$\pm$0.9 & 247.5$\pm$1.5 &   2   &-1.57$\pm$0.10 &     2& 170.0$\pm$10.0 &  35.0$\pm$10.0 \\ 
          &                &              &               &       &               &      &                &                \\ 
NGC\,1928 & 119.0$\pm$20.0 & 0.6$\pm$0.9  & 249.6$\pm$12.8&   4  &-1.30$\pm$0.15 & 4   &  85.0$\pm$10.0 &  45.0$\pm$10.0  \\  
NGC\,1939 & 144.2$\pm$16.0 & 0.8$\pm$0.9  & 258.8$\pm$7.4 &   4 &-2.00$\pm$0.15 & 4   & 130.0$\pm$10.0 &  35.0$\pm$10.0 \\\hline      
\end{tabular}

\noindent Ref.: (1)\citet{s92}; (2)\citet{getal06}; (3)\citet{shetal10}; (4) this work; (5)\citet{setal92}; 
(6)\citet{beetal2002}; (7)\citet{mucciarellietal2010}; (8)\citet{johnsonetal2006}; (9)\citet{wagnerkaiseretal2018}; 
(10)\citet{walker1992}; (11)\citet{matelunaetal2012}.
\end{table*}

\section{Conclusions}

With the aim of investigating the origin of the LMC GCs NGC\,1928 and 1939, we
carried out spectroscopic observations of giant stars located in their fields with 
the GMOS and the AAOmega+2dF spectrographs of the Gemini South and the Australian Astronomical Observatories, respectivey. The targets were selected
bearing in mind their positions along the red giant branch or red clump in  $HST$  cluster CMDs,
the only available photometric data set at the moment of preparing the observations. Some few candidates 
without $HST$  photometry were also selected.

The resulting high S/N spectra centred on the Ca\,II infrared triplet allowed us to measure accurate 
individual RVs for 11 and 15 stars in the fields of NGC\,1928 and 1939, respectively. The
RVs were obtained through cross-correlation of the observed spectra with template spectra.
We also measured equivalent widths of the three Ca\,II lines and derived individual metallicities 
([Fe/H]) for those stars with available photometry using a previous well-established calibration.
The accuracy in the individual [Fe/H] values ranges 0.1-0.3 dex.

By considering as membership probability criteria the position of the observed stars in the
cluster CMDs, and their position in the RV and metallicity distribution functions, we 
concluded that 7 and 9 observed stars are probable cluster members of NGC\,1928 and 1939, respectively.
The combined three criteria resulted to be a robust approach to assess the cluster
membership of the observed stars. From the adopted cluster members we estimated for the
first time accurate mean cluster RVs and metallicities. We found that NGC\,1928 is one
of the most-metal rich GCs ([Fe/H]=-1.3 dex), and NGC\,1939 is one of the most
metal-poor ones ([Fe/H]=-2.0 dex).

Both GCs are located in the innermost region of the LMC (deprojected distance <
1 kpc) and have RVs consistent with being part of the LMC disc. Therefore, we rule out
any possible origin but that in the same galaxy. Indeed, we computed the best solution for a
rotation disc that fully contains each GC, separately, and found that the
resulting circular velocities at the deprojected cluster distances very well match the rotation 
curves fitted from $HST$ and  $Gaia$ DR2 proper motions, respectively.

We extended our kinematics analysis to all the 15 LMC GCs by obtaining also
circular velocities. The outcomes show that most of the GCs share the LMC
rotation curve. Since they span the whole LMC GC metallicity range with no 
evidence of a metallicity gradient, we concluded that the LMC disc has existed since the
early epoch of the galaxy formation and has also experienced the abrupt chemical enrichment
seen in its GC populations in an interval of time of $\sim$ 3 Gyr. Four 
objects out of the fifteen GCs (NGC\,1835, 1898, 2210 and Reticulum) have
estimated circular velocities which notably depart from the LMC rotation curve. We think that
they are witnesses of having been stripped by the LMC from the SMC, an scenario predicted
from numerical simulations of the galaxy dynamical interactions and confirmed from
observation of field star populations.

\section*{Acknowledgements}

Based on observations obtained at the Gemini Observatory, which is operated by 
the Association of Universities for Research in Astronomy, Inc., under a cooperative
 agreement with the NSF on behalf of the Gemini partnership: the National Science
 Foundation (United States), the National Research Council (Canada), CONICYT (Chile),
 Ministerio de Ciencia, Tecnolog\'{i}a e Innovaci\'{o}n Productiva (Argentina),
 and Minist\'{e}rio da Ci\^{e}ncia, Tecnologia e Inova\c{c}\~{a}o (Brazil). 
We thank Dougal Mackey for providing us with the $HST$ photometric data base.
We thank the referee for the thorough reading of the manuscript and
timely suggestions to improve it. 




\bibliographystyle{mnras}

\input{paper.bbl}







\bsp	
\label{lastpage}
\end{document}